 \documentclass[a4paper,12pt]{article}
\pdfoutput=1
\usepackage{soul}
\usepackage[usenames,dvipsnames]{color}
\usepackage{jheppub}
\usepackage{hyperref}
\usepackage{amsmath, amssymb, slashed, epsf, color, graphicx, latexsym}
\begin{document}

	\title{Gravitostatic black hole Love numbers from Membrane Paradigm at large $D$ }
	\author{Subhajit Mazumdar}

	\affiliation{Okinawa Institute of Science and Technology, 1919-1 Tancha, Onna-son, Okinawa 904-0495, Japan}
 
	\emailAdd{subhajitmazumdar@gmail.com}

	\abstract{Black holes, like any other object - respond to external fields like background electric and gravitational fields. At lowest order, the response of a black hole to, (e.g.) a gravitational wave in the $l^{th}$ spherical harmonic mode is to develop a generalized `quadrupole moment' in the same symmetry channel. The induced moment of the black hole is proportional to the external forcing field. The constant of proportionality between the field and the response is a function of the frequency $\omega$ of the incident perturbation. These constants of proportionality are sometimes called `Love' numbers in the study of astrophysical black holes. Using the `Large $D$ Membrane Paradigm' formalism, we compute gravitostatic black hole `Love' numbers (at $\omega \to 0$ limit), in large $D$. This work opens the road of using Large $D$ Membrane Paradigm formalism to extend the computation of these black hole `Love' numbers for all values of $\omega$ in large $D$ - which we announce in this paper.}

\maketitle
\section{Introduction}
Black holes are well-understood and deeply mysterious solutions to Einstein’s equations. While exact analytical solutions exist for stationary black holes/ branes in simpler forms in all space-time dimensions, their apparent simplicity is deceptive, as their dynamic behaviour such as black hole collisions is highly complicated and hard to capture analytically. In string theory, the classical dynamics of black holes is central to the AdS/CFT correspondence. Classical black hole dynamics in Anti-
de Sitter (AdS) space help study strongly coupled field theories via the fascinating
AdS/CFT correspondence, revealing insights into hydrodynamics, entanglement, and
thermalization. A deeper understanding of classical black hole dynamics in complicated scenarios can provide profound insights into various areas of theoretical physics. However, due to complex nature of Einstein's equations it becomes very difficult to analytically control violent dynamical processes involving black holes. In this situation, the natural approach of a theoretical physicist of searching a small natural perturbative parameter to set up perturbative expansion of the problem, remains challenging as vacuum Einstein's equations are parameter-free.\\
\indent It has recently been demonstrated that the dynamics of a black hole in large space-time dimensions ($D$) is dual to the motion of a co-dimension one non-gravitational membrane propagating without back reaction in flat space (or generic space-times that solves Einstein equations).  Based on this observation several papers have been written in the subject Large $D$ Membrane Paradigm \cite{Bhattacharyya:2015dva, Dandekar:2016fvw, Dandekar:2016jrp, Dandekar:2017aiv}. The membrane hosts a stress tensor which is given in terms of the shape and velocity field \cite{Bhattacharyya:2016nhn}. The equations of motion for the membrane variables are generated by the requirement of membrane stress tensor conservation. This membrane--gravity correspondence (`Large $D$ Membrane Paradigm') was motivated by earlier observations \cite{Emparan:2013moa, Emparan:2013xia, Emparan:2013oza, Emparan:2014cia, Emparan:2014jca, Emparan:2014aba, Emparan:2015rva} by Emparan, Suzuki and Tanabe (EST) - where one can study the dynamics of the black holes in order by order in $\frac{1}{D}$. \\
\indent A quantity of interest in the study of the physics of materials is the polarizability of atoms and molecules. The question asked there is the following: What dipole moment does an atom develop when it is subjected to an electric field. The polarizability of atoms and molecules characterizes the response of these objects to electric fields, and is a key ingredient in the computation of material properties like the refractive index. In a similar manner in the study of black hole physics it is natural to ask the following question: how does a black hole respond when subjected to a gravitational perturbation? If the gravitational wave in question has angular momentum $l$ then the black hole clearly responds by developing some sort of `$l$ pole'. In the language of the membrane paradigm, the black hole membrane develops a shape (and, in general, velocity) deformation proportional to the $l^{th}$ spherical harmonic. The constant of proportionality between perturbation and response is an elastic constant analogous to atomic polarizability. In the special case that the incident radiation is in the scalar sector and is stationary, Kol and Smolkin (KS) \cite{Kol:2011vg} has defined and computed these elastic constants - so called `Love' numbers \footnote{Named after A.E.H. Love for his work \cite{love1909yielding} on gravitational tidal deformations. } - in arbitrary space-time dimensions $D$. For more works on black hole `Love' numbers see \cite{Damour:2009vw, Binnington:2009bb, Damour:2009va, Hadad:2024lsf}. \\
\indent We use the same stress tensor (in large $D$), which sources the radiation, reported in \cite{Bhattacharyya:2016nhn} to determine the gravitostatic (time independent) black hole `Love' numbers in Large $D$ Membrane Paradigm formalism. Our results of gravitostatic `Love' numbers, using the large $D$ Membrane Paradigm formalism, is proportional to the answer by KS in \cite{Kol:2011vg} and exactly agrees with the result of Cardoso, Gualtieri, and Moore (CGM) in \cite{Cardoso:2019vof} in the large $D$ limit - which can also be regarded as a consistency check of the large $D$ black hole stress tensor used in this paper. This computation of `Love' numbers presented in this paper opens the road to generalize `Love' numbers at nonzero $\omega$ (frequency dependent case) using our Large $D$ Membrane Paradigm formalism - these computations would be difficult to do in direct gravitational methods. In this paper we briefly announce our recently obtained results for frequency dependent black hole `Love' Numbers at large $D$ - a static limit ($\omega \to 0$) of which produces the static answer obtained in this paper.

 \section{Formulation and Calculation}\label{zeroomega}
 Black Hole `Love' numbers could be thought analogous to polarization of atom in the electric field. In the similar manner of measuring the electric polarizability, we could measure the black hole `Love' numbers. Let us suppose, in static case, $\phi$ be the field, in presence of the black hole, which is a combination of in-coming and out-going parts with leading radial dependence of $r^l$ and $\frac{1}{r^{l+D-3}}$ respectively, when the black hole is subjected to a gravitational wave in the $l^{th}$ spherical harmonic;  $\phi$ \footnote{In the expression of $\phi$, only the leading radial dependence has been highlighted. In general each of the in-coming and outgoing parts have higher order corrections.} at leading order, could be expressed as follows, 

 \begin{eqnarray}\label{lovedef}
     \phi = \hat{\phi} \left(r^l+ \frac{\lambda_l}{r^{l+D-3}}\right),
 \end{eqnarray}
 where $\lambda_l$ is some constant and  which is the measurement of the black hole `Love' number, where $\hat{\phi}$ is an overall constant. Basically, one needs to take ratios of the amplitude of the in-coming and the radiated wave (out-going) to determine the black hole `Love' number. See \cite{Kol:2011vg} for more details regarding the measurement of the black hole `Love' numbers.\\
 \indent We will be calculating Gravitostatic black hole `Love' numbers using `Large $D$ Membrane Paradigm formalism', in large $D$. We know the membrane hosts a stress tensor which is given in terms of its shape and velocity field. We will use the following space-time stress tensor $T_{\mu \nu}$, reported in  \cite{Bhattacharyya:2016nhn}, in the large $D$ limit, in our calculation. 
\begin{equation} \label{stt}
 T_{\mu\nu}=\frac{1}{16 \pi}\mathcal{K} u_\mu u_{\nu} 
\end{equation}

where $u_{\alpha}$ is a velocity field defined on the membrane and $\mathcal{K}$ is trace of the extrinsic curvature of the membrane.\\
\indent We will find the membrane equation that follow from the conservation of this stress tensor. We shine a stationary linearized wave onto a spherical black hole dual to a spherical non-gravitational membrane and compute the shape fluctuations of this membrane and thus the resultant radiation field. From our results we will read off the `Love numbers' as defined in \eqref{lovedef}. At large $D$ limit, we will see that our result is proportional to the answer obtained by KS \cite{Kol:2011vg} up to a factor, and exactly matches with the answer in a more recent work by CGM \cite{Cardoso:2019vof}.

\subsection{Linearized Einstein's equations}
We will now begin the first step of computation of the gravitostatic scalar black hole `Love' numbers. At first we will solve the linearized Einstein's equation to obtain the most general regular static solution as an expansion about flat space. We write the most general linear metric correction to the flat metric as,
\begin{eqnarray}\label{met11}
ds^2 &=& (-1+\epsilon h_{tt})dt^2 + \epsilon rh_{ta}dt d\theta^a + r^2(\Omega_{ab} +\epsilon h_{ab}) d\theta^ad\theta^b\\ \nonumber
\mbox{where, we assume}~~~~ h_{tt} &=& S_1\\ \nonumber
h_{ta} &=& V_a + \nabla_a S_2\\ \nonumber
h_{ab} &=& T_{ab} + \frac{\Omega_{ab}}{D-2}S_3 + \left(\nabla_a\nabla_b-\frac{\Omega_{ab}}{D-2}\nabla^2\right)S_4 +(\nabla_a \bar{V}_b + \nabla_b \bar{V}_a),
\end{eqnarray}
and it is clear that we fix $\nabla.V = 0$, $\nabla_a T^{ab} = 0 $ and $ T^a_a =0$. Please note that $\nabla_a$ is covariant derivative with respect to $\Omega_{ab}$ and we have fixed our gauge as $h_{r \alpha}=0$. It is understood that $t$ is the time index, $r$ is the radial index, $a$ runs over the angle indices and $\alpha$ runs over all type of indices. In the metric correction, an auxiliary small parameter $\epsilon$ is introduced to keep track of the linearity in our calculations. \\
\indent As we have made choice of our gauge condition as $h_{r \alpha}=0$, it is obvious that $E_{rt}$, $E_{ra}$ and $ E_{rr}$ would not contain any double derivative of $r$. Therefore, the equations they produce are not dynamical in nature, but these are constraint equations. Whereas $E_{tt}$, $E_{ta}$ and $E_{ab}$ produce dynamical equations. In terms of our metric specified in \eqref{met11}, these equations are given below in two sets, each containing three equations, namely constraint equations and dynamical equations, respectively.

\begin{eqnarray}
   2 R_{rt} &=& \frac{1}{r^2}\partial_r (r\nabla^2 S_2)\\
   2 R_{ra} &=& \partial_r(\nabla^ch_{ca})+\partial_r(\nabla_aS_1)-\frac{1}{r}\nabla_aS_1 - \partial_r\nabla_aS_3\\
   2 E_{rr} &=&\nonumber \frac{D-3}{r}\left(\partial_rS_3+\frac{S_3}{r}\right) +\frac{1}{r^2}\nabla^2S_3-\frac{1}{r^2}\nabla^c\nabla^dh_{cd}-\frac{1}{r^2}\nabla^2S_1-\frac{D-2}{r}\partial_rS_1\\
\end{eqnarray}

\begin{eqnarray}\label{DynEqn}
    2 R_{tt} &=& -\partial^2_rS_1 -\frac{D-2}{r}\partial_rS_1 -\frac{1}{r^2}\nabla^2S_1\\
    2 R_{ta} &=& -\partial^2_r(r h_{ta}) -\frac{D-4}{r}\partial_r(r h_{ta})+\frac{D-3}{r}h_{ta}-\frac{1}{r}\nabla^2h_{ta}+\frac{1}{r}\nabla_a(\nabla^2S_2)\\
    \nonumber 2 R_{ab} &=& -\partial^2_r(r^2 h_{ab})-\frac{D-6}{r^2}\partial_r(r^2h_{ab})+2(D-4)h_{ab}-\nabla^2h_{ab}\\\nonumber &-&\left(\frac{1}{r}\partial_r(r^2S_3)-r\partial_rS_1\right)\Omega_{ab} -\nabla_a\nabla_b(S_3-S_1)+\nabla_a\nabla^ch_{cb}+\nabla_b\nabla^ch_{ca}\\
\end{eqnarray}

\subsection{Calculation of membrane data }
 We will now calculate the response of a black hole in presence of incident gravitational wave, in large $D$.  $K_{\alpha \beta}$ is the extrinsic curvature tensor which is defined from the normal $n_\mu$ to the membrane surface, i.e. $K_{\alpha \beta}= \nabla_\alpha n_\beta $.  The covariant derivatives are defined for the metric $\eta_{\mu \nu } + \epsilon h_{\mu \nu}$ defined in \eqref{met11}, up to the linear order in $\epsilon$. \\
\indent The shape and velocity field on the membrane are defined as follows. For our system, without any loss of generality we assume that the membrane is at, $r= 1+ \epsilon \delta r\left(\{\theta^a\}\right)$ and the velocity on the membrane is  $u_\alpha dx^\alpha = -dt + \epsilon \delta u_A dx^A$ \footnote{Where A runs over all space-time indices but time.}. The shape and velocity fluctuations are denoted by $\delta r$ and $\delta u_A$, respectively. Then the induced metric on the membrane is given by,
\begin{equation}\label{imet}
    ds^2 = (-1+\epsilon h_{tt})dt^2+dr^2+ 2\epsilon r h_{ta} dt d\theta^a + r^2(\Omega_{ab}+ \epsilon h_{ab}) d\theta^a d\theta^b.
\end{equation}
\indent On the membrane, the components of the extrinsic curvatures are given by,
\begin{eqnarray}\label{ec} \nonumber
    K_{tt} &=& \frac{\epsilon}{2} \partial_r h_{tt}\\\nonumber
    K_{ta} &=& \frac{\epsilon}{2}(\partial_rh_{ta} + h_{ta})\\ 
    K_{ab} &=& \Omega_{ab} + \epsilon \delta r \Omega_{ab} -\epsilon\nabla_a\nabla_b\delta r + \frac{\epsilon}{2} \partial_r h_{ab} + r_{0} \epsilon h_{ab} 
\end{eqnarray}
and trace of the extrinsic curvature is evaluated to,
\begin{equation}\label{tec}
    \mathcal{K} = D -\epsilon \nabla^2\delta r - \epsilon (D)\delta r -\frac{\epsilon}{2}\partial_rh_{tt}+\frac{\epsilon}{2}\partial_r h_{ab}\Omega^{ab}.
\end{equation}
\indent The components of the world-volume stress tensor \footnote{World-volume stress tensor is obtained by multiplying $\delta (r-1-\epsilon \delta r )$ with space-time stress tensor.}, obtained from \eqref{stt}, are given by,

 \begin{eqnarray}\label{sttt}\nonumber
 T_{tt}&=&\frac{1}{16 \pi}\left(\mathcal{K}+K_{tt}-\epsilon \mathcal{K} h_{tt}\right)\delta (r-1-\epsilon \delta r )\\\nonumber
 T_{ta}&=&\frac{1}{16 \pi}\left(D(\epsilon h_{ta}-\epsilon \delta u_{a})+K_{ta}-\epsilon h_{ta} \mathcal{K}\right)\delta (r-1-\epsilon \delta r )\\\nonumber
T_{ab}&=&\frac{1}{16 \pi}\Bigg(D\left(\Omega_{ab}+\epsilon (2 \delta r \Omega_{ab}+h_{ab})+\frac{\epsilon}{2}h_{tt}\Omega_{ab}\right)\\
&+& K_{ab}-\mathcal{K}\Omega_{ab}-\epsilon (2\delta r \Omega_{ab} \mathcal{K}+h_{ab}\mathcal{K})\Bigg)\delta (r-1-\epsilon \delta r )).
\end{eqnarray}
\indent It can be shown that the membrane obeys the following equation of motion, at large $D$ - obtained from the stress tensor conservation equation projected orthogonal to the velocity. To gain a better understanding of the genesis of the following equation, see \cite{Dandekar:2017aiv}.

\begin{equation}\label{ecstt}
  \mathcal{K} =D \frac{1}{\sqrt{-g_{tt}}}
\end{equation}

\indent We substitute the value of $\mathcal{K}$ from \eqref{tec} and $g_{tt}$ from \eqref{imet} and solve equation \eqref{ecstt} to obtain the shape fluctuation, at large $D$, as,
\begin{equation} \label{fluc}
 \delta r= \frac{h_{tt}}{2(l-1)}=\frac{S_1}{2(l-1)}.
\end{equation}
\indent Now, this wiggles or shape fluctuations on the membrane, due to in-coming gravitational wave, will induce radiation.
\subsection{Calculation of gravitostatic black hole `Love' Numbers}
We can see from \eqref{DynEqn}, that the decoupled differential equation for the scalar mode comes from the $tt$ component of the linearized Einstein's equations, for our specific choice of gauge condition. We will first solve it switching the source part (containing stress tensor) to zero in order to find the in-coming solution. Below we express the $tt$ component of Einstein's equation with zero source ($R_{tt}=0$). \\
\indent As we know that,
\begin{eqnarray}
&&2 R_{tt} = -\partial^2_r h_{tt} - \frac{D-2}{r} \partial_r h_{tt} - \frac{1}{r^2}\nabla^2 h_{tt},
\end{eqnarray}
then $R_{tt}=0$ suggests,
\begin{equation}
     -\partial^2_r h_{tt} - \frac{D-2}{r} \partial_r h_{tt} - \frac{1}{r^2}\nabla^2 h_{tt}=0.
\end{equation}
\indent Using \eqref{met11} we get the following equation,
\begin{equation}\label{dectt}
\partial^2_r S_1 + \frac{D-2}{r} \partial_r S_1 + \frac{1}{r^2}\nabla^2 S_1 = 0.
\end{equation}
\indent We now input our ansatz for the solution as $S_1 = R_1(r) \Theta_1(\theta^a)$ in \eqref{dectt} for separation of the radial and the angular part, which gives an ordinary differential equation for the radial part, as follows, which is readily solvable.
\begin{equation}
   \frac{d^2 R_1}{dr^2} + \frac{D-2}{r} \frac{dR_1}{dr} - \frac{l(D+l-3)}{r^2}R_1 = 0 
\end{equation}
\indent The general solution for $R_1$ is given by,
\begin{equation}
 R_1 = A_1 r^l + \frac{A_2}{r^{l+D-3}}.
\end{equation}
\indent Regularity condition suggests $A_2=0$. Therefore, we get $S_1$ as follows,
\begin{equation}\label{ins}
    S_1 = \sum_{l=0}^\infty A_{lm} r^l Y_{lm}(\theta^a).
\end{equation}
\indent We have found that there is one scalar, one vector and one tensor mode which are regular at $r=0$. On the membrane we find one scalar to be $h_{tt}$ and $h_{ta}$ as one vector mode. For our computation we only need the equation for the scalar. We know that the Einstein's equation, in $D$ space-time dimensions, with source, for the scalar mode is given by,
\begin{eqnarray}\label{einn}
 R_{tt}= 8 \pi(T_{tt}+\frac{1}{2(1-\frac{D}{2})}g_{tt} (g^{\alpha \beta} T_{\alpha \beta})\delta(r-1-\epsilon \delta r).
\end{eqnarray}
\indent The R.H.S. of \eqref{einn} could be simplified using stress tensor components from \eqref{sttt}. At large $D$ limit and for $l \neq 0$ modes we get, 
\begin{eqnarray}\label{einnD}
 R_{tt} = -\epsilon \frac{D}{2} \partial_r \delta(r-1)\delta r=S_{lm}Y^{lm}.
\end{eqnarray}
\indent From \eqref{DynEqn} we know,
\begin{eqnarray}\label{lein}
R_{tt}=-\frac{\epsilon}{2} \left( \partial^2_r h_{tt} + \frac{D-2}{r} \partial_r h_{tt} + \frac{1}{r^2}\nabla^2 h_{tt}\right).
\end{eqnarray}
\indent Putting $h_{tt} = \phi_l(r) Y_{lm}(\theta^a)$ in \eqref{lein} and removing the angular part, we get
\begin{eqnarray} \label{pheq1}
\frac{\epsilon}{2}\left( \frac{d^2 \phi_l}{dr^2} + \frac{D-2}{r} \frac{d \phi_l}{dr} - \frac{l(D+l-3)}{r^2}\phi_l\right) =- S_{lm}.
\end{eqnarray}
\indent The operator on $\phi_l$ in \eqref{pheq1} could be recognized as $\nabla^2$ in $D$ dimensions, therefore \eqref{pheq1} could be re-written as,
\begin{equation} \label{pheq}
    \nabla^2 \phi =  -S_{lm}.
\end{equation}
\indent To solve \eqref{pheq} we will start with the following equation and solve it by perturbation technique (separating the zero mode).
\begin{equation}\label{Eqn}
\nabla^2 \phi = D \delta(r-(1+\epsilon \delta r))
\end{equation}
\indent We assume that the following in-coming wave ($\Phi_{in}$) and out-going wave ($\Phi_{out}$) are the solutions (ansatz) \footnote{It can be easily observed that the zeroth order solutions are manifestly consistent with \eqref{Eqn} - with its R.H.S. switched to 0, and they match at the horizon - assumed to be at $r=1$.} of \eqref{Eqn}.
\begin{eqnarray}\nonumber
    \Phi_{out}&=&-\frac{1}{r^{D-3}}+\epsilon ~ \frac{b_{lm} Y^{lm}}{r^{l+D-3}}\\
    \Phi_{in}&=&-1+\epsilon ~ a_{lm} Y^{lm}r^{l}
\end{eqnarray}
\indent We are going to solve a second order differential equation, and therefore, we need two specify boundary conditions. We use the first boundary condition as regularity at the horizon which translates that both of the solutions have to agree at $r=1+\epsilon \delta r$. 
\indent We therefore demand,
\begin{eqnarray}
\Phi_{in}|_{r=1+ \epsilon \delta r}=\Phi_{out}|_{r=1+ \epsilon \delta r}.
\end{eqnarray}
\indent This gives us,
\begin{eqnarray}\label{fbc1}
-1+\epsilon a_{lm}Y^{lm}(1+\epsilon \delta r)^{l}=-\frac{1}{(1+\epsilon \delta r)^{D-3}}&+&\epsilon \frac{b_{lm}Y^{lm}}{(1+\epsilon \delta r)^{(l+D-3)}}.
\end{eqnarray}
 \indent \eqref{fbc1} simplifies to,
\begin{equation}
    -(D-3) \delta r +a_{lm}Y^{lm}-b_{lm}Y^{lm}=0.
\end{equation}
\indent Second boundary condition of pillbox (discontinuity at horizon) demands,
\begin{equation}\label{sbc2}
\left(\nabla_{r}\Phi_{out}-\nabla_{r}\Phi_{in}\right)|_{r=1+ \epsilon \delta r}=D
\end{equation}
\indent After simplifications, this condition gives us,
\begin{eqnarray}
-(D-3) (D-2) \delta r-(l+D-3)b_{lm}Y^{lm}-l a_{lm}Y^{lm}=0.
\end{eqnarray}
\indent We will only focus on the part where $l \neq 0$ for our necessary solution which is,
\begin{equation}
b_{lm}Y^{lm}=-\frac{ (D-3)(D+l-2)\delta r}{(2l+D-3)}.
\end{equation}
\indent As per \eqref{lovedef} the black hole `Love' Number  will be obtained by the ratio of the amplitudes of out-going wave  and the in-coming scalar mode $S_1$ which is given in \eqref{ins} and using the answer for the shape fluctuation in \eqref{fluc}. We finally obtain the answer for the gravitostatic black hole `Love' number $\lambda_l$ \footnote{In this calculation, without any loss of generality, the Schwarzchild radius of the black hole is assumed to be at $r=1$. If the Schwarzchild radius is assumed to be at $r=r_0$ , it can be shown from dimensional analysis that the answer of the black hole `Love' Number in \eqref{final love number} should be multiplied by $r_0^{2l+D-3}$.}, at large $D$, as follows,
\begin{eqnarray}\label{final love number}\nonumber
\lambda_{l}= b_{lm}Y^{lm} \bigg /S_1  &=& -\frac{ (D-3)(D+l-2)\delta r}{(2l+D-3)}\bigg/ S_1\\\nonumber
&=&  -\frac{ (D-3)(D+l-2)}{(2l+D-3)} \frac{S_1}{2(l-1)} \bigg/ S_1\\
&\sim& -\frac{D}{2(l-1)} ~~~~~~~~~~(\text{at leading order in $D$}).
\end{eqnarray}
\indent KS \cite{Kol:2011vg} obtained the following answer for the gravitostatic black hole `Love' number which we denote by $\lambda_{KS}$ \footnote{In the expression of KS answer  $\lambda_{KS}$ in \eqref{ksa} the Schwarzchild radius of the black hole is assumed to be at $r=1$.}.

\begin{equation}\label{ksa}
    \lambda_{KS}= \frac{\Gamma (\frac{l}{D-3}) \Gamma(\frac{l}{D-3}+2)}{\Gamma (\frac{l}{D-3}+\frac{1}{2}) \Gamma(\frac{l}{D-3}+\frac{3}{2})} \tan \left(\pi \frac{l}{D-3}\right) \left(\frac{1}{4}\right)^{\frac{2l}{D-3}+1}
\end{equation}
\indent CGM \cite{Cardoso:2019vof} observed that their answer of the gravitostatic `Love' number of black holes denoted by $\lambda_{CGM}$ \footnote{Similar to KS answer in \eqref{ksa}, in the expression of $\lambda_{CGM}$ in \eqref{cgma} the Schwarzchild radius of the black hole is also assumed to be at $r=1$. } - is proportional with KS answer $\lambda_{KS}$, with an overall proportionality factor shown below,
\begin{equation}\label{cgma}
     \lambda_{CGM} = \bigg(-\frac{(D-2+l)}{l-1}\bigg) \lambda_{KS}.
\end{equation}
\indent If we now calculate the CGM answer, in large $D$, we get, the following,
 \begin{eqnarray}\nonumber \label{cgmans}
     \lambda_{CGM} &=& \bigg(-\frac{(D-2+l)}{l-1}\bigg) \lambda_{KS}\\\nonumber
     &=&\bigg(-\frac{(D-2+l)}{l-1}\bigg) \frac{\Gamma (\frac{l}{D-3}) \Gamma(\frac{l}{D-3}+2)}{\Gamma (\frac{l}{D-3}+\frac{1}{2}) \Gamma(\frac{l}{D-3}+\frac{3}{2})} \tan \left(\pi \frac{l}{D-3}\right) \left(\frac{1}{4}\right)^{\frac{2l}{D-3}+1}\\
     &\sim& -\frac{D}{2(l-1)} ~~~~~~~~~~(\text{at leading order in $D$}).
 \end{eqnarray} 
\indent On comparison of our answer in \eqref{final love number} and CGM answer at large $D$ in \eqref{cgmans}, we can easily see that they matches exactly.
 
 \section{Announcement of results in frequency dependent case}\label{secann}
 In this paper, we would like to take the opportunity to announce our very recently obtained results of black hole `Love' Number in the time dependent case, at large $D$. A dedicated paper discussing the determination of time ($t$)/ frequency ($\omega$) dependent black hole `Love' Number $ \lambda_l \left( \omega \right)$ is currently under preparation and we just report our result below. A follow up paper containing the full analysis will appear very soon. In that computation, we had to fix our gauge differently - similar to \cite{Bhattacharyya:2016nhn}, for technical reasons, as the radiation due to shape fluctuations, for incident frequency dependent gravitational wave, was calculated following results in that paper. It is very satisfying to notice that if $\omega \rightarrow 0$ limit is taken in the time-dependent answer mentioned below, we recover our above-mentioned answer in \eqref{final love number} for the static case - which is a consistency test of our time-dependent result where the black hole `Love' number is a function of finite frequency $\omega$.

 \begin{eqnarray}
     \lambda_l \left( \omega \right)= \frac{D  \left(\frac{l}{2}-i \omega \right)}{ (\omega -\omega_1) (\omega -\omega_2)} e^{-i \omega t}
 \end{eqnarray}
 \indent where, $\omega_1$ and $\omega_2$ are the two light quasinormal mode frequencies, namely,
 \begin{equation}
 \omega_1=\sqrt{l-1}-i (l-1)
 \end{equation}
 \indent and
 \begin{equation}
 \omega_2=-\sqrt{l-1}-i (l-1).
 \end{equation}
\section{Discussion and Future Directions}
In this paper we have determined the gravitostatic polarizability constants (`Love' Numbers) of a spherical black hole due to static scalar gravitational perturbation using `Large $D$ membrane paradigm' formulation, at large $D$. We have found that it is proportional with the answer of KS \cite{Kol:2011vg} and exactly matches with the answer of CGM \cite{Cardoso:2019vof}. We have derived in our earlier work \cite{Dandekar:2017aiv} an improved stress tensor which has a finite $D$ completion; it captures probe membrane dynamics at finite $D$ while returning results dual to a black hole at large $D$. We plan to use this improved stress tensor to calculate the gravitostatic black hole `Love' number and see how much it agrees to the literature, at finite $D$. It would be wonderful if we also manage to recover the exact overall normalization factor (in finite $D$) observed by CGM \cite{Cardoso:2019vof} w.r.t. KS answer. It would also be interesting to understand the physics of this factor.\\ 
\indent As we mentioned in section \ref{secann}, we are going to send out a detailed paper which extends the computation of these black hole `Love' numbers using the Large $D$ Membrane Paradigm, for all values of $\omega$ in large $D$, which would be difficult to do directly in Gravity. We have announced our answer of frequency dependent black hole `Love' numbers in section \ref{secann} and we view the exact matching of this answer at $\omega \to 0$ limit with our black hole `Love' number answer obtained in this paper for static case to be non-trivial. In other words, the Large $D$ Membrane Paradigm gives us new results in addition to reproducing known results, demonstrating the power of this tool. We would like to further extend these computations of `Love' Numbers for rotating and charged black holes/ branes moving in flat or more generic backgrounds, such as $dS$ or $AdS$. \\
\indent Recently, Large $D$ Membrane equations with cosmological constants \cite{Bhattacharyya:2017hpj, Bhattacharyya:2018szu} and for higher derivative gravity theories \cite{Saha:2018elg, Dandekar:2019hyc} were derived. It would be nice to determine black hole `Love' numbers for higher derivative gravity theories using our powerful tool `Large $D$ Membrane Paradigm' formalism. \\
\indent We hope to return to these above-mentioned calculations in the future.

 \section*{Acknowlegdements}
  The author would like to especially thank Shiraz Minwalla, who has always been in the advisory capacity of this project, and the author is indebted to him for several extremely useful discussions. The author thanks Yogesh Dandekar, Suman Kundu and Arunabha Saha for collaborations during initial stages of this work and related projects. The author was benefited from extremely useful discussions with Roberto Emparan, Barak Kol, Sangmin Lee and Yasha Neiman, and thanks all of them. The author sincerely acknowledges the hospitality of several institutes and organizers for hospitality during conferences, workshops, and official visits during the span while this work was in progress. This work was done in several phases while the author was affiliated with different institutes. The author acknowledges the support of these institutes and grants that supported this work at different stages. The work was supported, in part, by a UGC/ISF Indo Israel grant, the Infosys Endowment for Research into the Quantum Structure of Spacetime, “Quantum Universe” I-CORE program of the Israeli Planning and Budgeting Committee and the National Research Foundation of Korea grant NRF-2019R1A2C2084608. This work was mainly supported by the Quantum Gravity Unit of the Okinawa Institute of Science and Technology Graduate University (OIST).
 
\bibliographystyle{JHEP}
\bibliography{ssbib}

\end{document}